\documentclass{article}
\usepackage{graphicx}

\title{Instabilities in Granular Flows}
\author{
{\bf Shaun Hendy}\\ ~\\
{\em IRL Applied Maths,} \\
{\em Industrial Research Ltd,} \\
{\em PO Box 31-310, Lower Hutt,} \\
{\em New Zealand} \\
{\em s.hendy@irl.cri.nz} 
}

\begin{document}
\newcommand{\dnorm}[1]{\mbox{$\left|\left| {#1} \right| \right|$}}
\maketitle
\section{Introduction}

The flow and handling of granular materials is of major importance to many industries. Yet
despite efforts over several decades, the modeling of such flows has achieved only modest 
success. Dense gravity-driven flows in hoppers have been often modeled as elastic-plastic 
continua, for example. In this picture, the granular material flows as a plastic with a 
frictional yield condition, and deforms as an elastic solid otherwise. This model has been 
used to analyze mass flows, where the material is flowing throughout the hopper, but has failed 
to provide adequate agreement with experiment in the prediction of quantities such as discharge
rate, for example \cite{thorpe92}. Despite such shortcomings, Jenike's \cite{jenike} construction 
of steady-state incompressible rigid-plastic radial solutions in hoppers with simple geometries 
has been of considerable importance in hopper design \cite{nedderman92}. These are solutions 
for quasistatic flow (inertial effects are neglected) where grains travel radially in conical 
or wedge-shaped hoppers. Only recently have numerical solutions of steady-state quasistatic 
flows in more complicated geometries been produced \cite{gremaud99}.

However, it is now clear that there are serious mathematical difficulties with the equations 
for time-dependent incompressible rigid-plastic flow (IRPF). In many instances, the equations 
for such flows have been shown to be ill-posed i.e. they possess instabilities with
arbitrarily short wavelengths (Schaeffer \cite{schaeffer87}, Valanis and Peters \cite{peters96}). 
Hence, it is problematic to interpret the steady-state rigid-plastic flow equations as long-time 
solutions of the time-dependent equations. Additionally, both the steady-state and time-dependent 
equations have physical shortcomings. There are several features of hopper flows where the particle 
size is important, indicating that it may not be appropriate to model such dense granular flows 
as continua.

\begin{figure}
\hspace{-1.2cm}\includegraphics[scale=0.7]{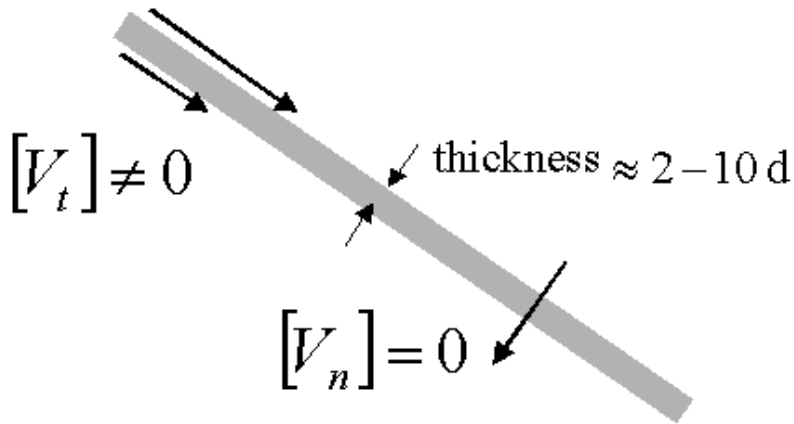}
\caption{Schematic illustration of shear banding. In granular materials the band is typically 
several grain diameters (d) thick.}
\label{figureI1}
\end{figure}

One flow feature that involves the particle size is that of shear-banding. Shear-banding is an 
extremely interesting phenomena that occurs in plastically deforming materials, including granular 
materials. At high strains the deformation of the material becomes localized in thin bands of 
shearing. This leads to a jump in velocity across the band as illustrated in figure~\ref{figureI1}. 
In granular flows, the thickness of this band is typically only several grains thick, yet the 
length can be comparable to the size of the container. Thus shear-banding in granular flows seems to 
be a flow feature that spans both the kinetic and the hydrodynamic regimes. In this sense, 
shear-banding in granular materials differs, for example, from the discontinuities encountered 
in compressible fluid flows where shocks are smoothed by hydrodynamic effects (e.g. viscosity). 
The full theory of granular flow ought to predict a shear band thickness of zero in the 
hydrodynamic limit. 

It is probable that a complete description of shear-banding in granular flows is not within the 
scope of a continuum theory. Nonetheless, attempts have been made to describe shear-banding in 
continuum theories in some average sense (\cite{muhlhaus96}). Indeed, continuum plastic flow 
theories have been found to exhibit shear-banding type behavior and the development of shear-bands 
seems to be closely related to the development of ill-posedness in such flows \cite{schaeffer92b}. 
One approach to deal with ill-posedness is to model the granular material as a Cosserat continua. 
Cosserat-type continuum theories attempt to model materials with internal structure by the inclusion 
of extra terms motivated by physics at the granular scale. These extra terms can damp the instabilities 
that lead to ill-posedness and can be used to predict shear-band thickness (Muhlhaus \cite{muhlhaus96}). 
Including such damping terms which act at a granular length-scale and `thickening' the shear-bands 
is a practical approach to studying granular flows using continuum equations. Indeed, such 
an approach is analogous to including Reynolds stresses in the Navier-Stokes equations to model 
turbulence.

The purpose of this paper is to study a set of equations for granular flow, which are a simple 
generalization of the IRPF equations, and which display shear-banding instabilities similar to those
that arise in the Critical State theory (CST) of granular flow \cite{schaeffer90}. These equations appear 
to be more regular than the full IRPF equations and are somewhat simpler than CST-motivated equations. We 
do not explicitly introduce a granular length scale here; we wish to study the development of shear-banding 
from a hydrodynamic viewpoint. In the first section we state the usual equations for IRPF. In the next 
section , we review the linear stability analysis of these equations and show that they are linearly 
ill-posed in two dimensions. In section four, we introduce a new set of equations which are a singularly 
perturbed form of the IRPF equations, and in the next section we present numerical solutions of these 
equations which demonstrate shear-banding. Finally, we look at the linear stability of these equations to 
provide a qualitative description of the shear-banding process.

\section{Equations for an incompressible rigid-plastic flow}

We consider the flow under gravity of an incompressible granular material in a wedge-shaped hopper 
under plane strain. We model the flow of this material as a continuum rigid-plastic flow.  The 
equations for such a flow consist of the incompressibility condition, 
\begin{equation}
\label{I0}
\nabla_i \, u^i = 0
\end{equation}
where $u^i$ are the components of the velocity field, and the momentum equations:
\begin{equation}
\label{I1}
\rho \left( {\partial u^i \over \partial t}+ u^j \nabla_j {u^i} \right) + \nabla_j \sigma^{ij}  
=  \rho g^i,
\end{equation}
where $\sigma_{ij}$ is the symmetric stress tensor, and $g^i$ is the acceleration due to 
gravity. Note that we define $\sigma$ to be positive when forces are compressive.

Plastic deformation is assumed to occur everywhere in the hopper i.e. the material 
is at plastic yield throughout the hopper. We will use a flow rule based on a 
Drucker-Prager type yield condition \cite{drucker}. Specifically, in terms of the 
principal stresses $\sigma_i$, this condition is written as
\begin{equation}
\label{I2}
(\sigma_1 - \sigma)^2 + (\sigma_2 - \sigma)^2 + (\sigma_3 - \sigma)^2 \leq k^2 \sigma^2
\end{equation}
where $k=\sqrt{2} \sin \varphi$, $\varphi$ is the internal angle of friction of the material 
and $\sigma = {1 \over 3} (\sigma_1 + \sigma_2 + \sigma_3)$ is the average trace of the stress 
tensor (we will refer to $\sigma$ as the average stress). If this inequality is satisfied exactly 
then the material is deforming plastically. Under plane strain $\sigma_2 = \sigma = {1 \over 2} 
(\sigma_1 + \sigma_3)$, so in this case we may consider a strictly two-dimensional yield condition:
\begin{equation}
\label{I2.5}
(\sigma_1 - \sigma)^2 + (\sigma_3 - \sigma)^2 = 2 \sigma^2 \sin^2 \varphi
\end{equation}   
   
We now assume a flow rule of the form 
\begin{equation}
\label{I4}
\sigma_{ij} = \sigma \delta_{ij} + \mu V_{ij},
\end{equation} 
where $V_{ij}=(\nabla_i u_j + \nabla_j u_i)/2$ and $\mu$ is some, as yet unspecified, scalar function 
of the normal stress and strain rates. If we compare the flow rule (\ref{I4}) to the yield condition 
(\ref{I2}), then we see  such a flow will satisfy the yield condition exactly if we choose the 
function $\mu$ to be
\begin{equation}
\label{I5}
\mu = { k \sigma \over \dnorm{V}}. 
\end{equation}     
where $\dnorm{V}=\sqrt{V_{ij} V^{ij}}$. 

Equations (\ref{I0}), (\ref{I1}), (\ref{I4}) and (\ref{I5}) form a closed system for incompressible 
rigid-plastic flow in plane strain. For granular flows in hoppers, these equations are only valid for 
so-called mass flows where the material is flowing throughout the hopper. When the hopper is not 
sufficiently steep funnel flows can develop where material flows down a central funnel leaving a 
stagnant region adjacent to the walls. Indeed, radial solutions have been used to study the transition 
between mass and funnel flow which is thought to occur when the rigid-plastic equations become singular 
as the rate of deformation vanishes \cite{drescher}. We will restrict our attention to mass flows where 
rigid-plastic flow occurs everywhere in a given domain. 

On each wall of a hopper, the direction of flow is tangent to the wall and the stresses must satisfy 
a Coulomb friction condition. The Coulomb boundary condition in two dimensions can be written as 
follows:
\begin{equation}
\label{I6}
\sigma_{ij} n^i t^j  = \tan \varphi_w \,\, \sigma_{ij} n^i n^j,
\end{equation}
where $\vec{n}$ is a outward-pointing normal vector to the wall, $\vec{t}$ is a tangent vector to the 
wall pointing in the opposite direction to that of the flow and $\varphi_w$ is the angle of wall friction.

Combining equations (\ref{I0}), (\ref{I1}), (\ref{I4}) and (\ref{I5}) together we obtain the equations:
\begin{eqnarray}
\label{I7}
{\partial u^i \over \partial t}+ u^j \nabla_j {u^i} &=& g^i - \nabla_j \left( p \delta^{ij} 
- k p A^{ij}  \right), \\
\label{I7.1}
\nabla_i u^i &=& 0,
\end{eqnarray}
where \begin{eqnarray}
\nonumber p &=& \sigma/\rho,\\
\nonumber A_{ij} & = & V_{ij}/ \dnorm{V}.
\end{eqnarray}
These equations are strictly hyperbolic in the case of plane strain (equation (\ref{I4})). 

\section{Ill-posedness of the incompressible rigid-plastic flow equations}

The equations (\ref{I7}) and (\ref{I7.1}) have been shown to be linearly ill-posed in certain geometries 
and for certain parameter values \cite{schaeffer87}. Specifically, the linearized equations of motion that 
describe the propagation of a small disturbance in the flow, possess unstable plane-wave solutions in the
short wavelength limit. It has been suggested that these instabilities are related to the formation 
of shear bands in the flowing granular material. In this section, we will consider the linearized equations 
of motion for a plane-wave disturbances to look for ill-posedness. This ill-posedness in the IRPF equations 
was originally noted by Schaeffer \cite{schaeffer87} for fully three-dimensional flows. Here we will work 
only in 2D, but with the full linearized equations of motion for a small disturbance. This will serve to 
provide a comparison with work in the following sections.

We will begin by writing the two-dimensional rigid-plastic equations (\ref{I7} and \ref{I7.1}) in 
non-dimensional form as follows:
\begin{equation}
\hat{u}^i = u^i / u_0, \,\,\, \hat{p} = p/ g L, \,\,\, \hat{x}^i = x^i/L, \,\,\, \hat{t} = t \, u_0/L,
\end{equation}
where $u_0$ is some characteristic velocity and $L$ is some characteristic length-scale of the problem. 
The equations for the rigid-plastic flow then become 
\begin{eqnarray}
\label{L0}
{\partial \hat{u}^i \over \partial \hat{t}}+ \hat{u}^j \hat{\nabla}_j {u^i} &=& {1 \over Fr^2}
\left( g^i/g - \hat{\nabla_j} \left[ \hat{P} \delta_{ij} - k \hat{p} A_{ij} \right] \right), \\
\label{L0.1}
\hat{\nabla}_i \hat{u}^i &=& 0.
\end{eqnarray}
where $Fr=u_0 / \sqrt{g L}$ is the Froude number. At this stage will drop the $\hat{\,}$ 
notation and assume that we are dealing with dimensionless quantities unless otherwise specified. 

The linearized equations of motion for a small disturbance $(\delta \vec{u}, \delta p)$ propagating on 
a smooth background solution $(\vec{u},p)$ to equations (\ref{L0}) and (\ref{L0.1}) can be shown to be:
\begin{eqnarray}
\label{L1}
{\partial \delta u^i \over \partial t}+ u^j \nabla_j {\delta u^i} + \delta u^j \nabla_j {u^i} &=& 
{1 \over Fr^2} \nabla_j \left( k \left( \delta p \, A^{ij} + p \, \delta A^{ij} \right) -  
\delta p \, \delta^{ij}
\right), \\
\label{L1.1}
\nabla \cdot \delta \vec{u} &=& 0,
\end{eqnarray}
where
\begin{equation}
\nonumber \delta A_{ij} = \left(\delta_{i(k} \delta_{l)j} - A_{ij} A_{kl}\right) 
{\nabla^k \delta u^l \over \dnorm{V}}.
\end{equation}

We consider plane-wave disturbances $(\delta \vec{u}, \delta p) = \exp{(\lambda t + i \xi \cdot x)} 
(\vec{a}, \alpha)$ propagating with wavevector $\xi$. In general $\alpha$ and $\vec{a}$ will be 
complex quantities. From the linearized equations, we obtain the following relations for $\lambda$, 
$\vec{a}$ and $\alpha$:
\begin{eqnarray}
\label{L2}
\lambda a_i &=& B_i \alpha + C_{ij} a^{j}, \\
\label{L2.1}
\xi \cdot \vec{a} &=& 0,
\end{eqnarray}
where 
\begin{equation}
\label{L2.2}
B_i = {1 \over Fr^2} \left( k \nabla^j A_{ij}+ i \left(k A_{ij}-\delta_{ij} \right) \xi^j \right),
\end{equation}
\begin{equation}
\label{L2.3}
C_{ij} = -\left( \nabla_j u_i + i (\xi \cdot u) \delta_{ij} \right)+{\xi^l \over Fr^2} \left(- \mu M_{ijkl} \xi^k + i 
\nabla^k \left( \mu M_{ijkl} \right) \right),
\end{equation}
and
\[
M_{ijkl}=\left(\delta_{i(j} \delta_{l)k}-A_{il} A_{jk} \right).
\]
One can solve (\ref{L2}) and (\ref{L2.1}) for $\lambda$ for every wavevector $\xi$. The real part of $\lambda$ 
determines the growth or decay of the plane-wave disturbance with wavevector $\xi$, and the imaginary part 
determines the propagation of the disturbance. If the real part of $\lambda$ is positive for any $\xi$, we refer 
to this mode as {\it linearly unstable}, as this mode will grow rapidly in time. If, for a given background 
solution, (\ref{L2}) and (\ref{L2.1}) have unstable modes, then this solution will be termed {\it linearly 
unstable}. If, for a given background solution, there are unstable modes with arbitrarily short wavelengths, and 
if this background solution is a unique solution of the IPRF equations, then we will call these equations and 
the solution {\it linearly ill-posed}.  
                                       
Using the condition $\xi \cdot \vec{a} = 0$ we can eliminate $\alpha$ from the equation for $\lambda$:
\begin{eqnarray}
\label{L3}
\lambda a^i &=& D_{ij} a^j, \\
\label{L3.1}
D_{ij} &=& \left(\delta_{ik}- {B_i \xi^k \over \vec{B} \cdot \vec{\xi}} \right) C_{kj}, \\  
\label{L3.2}
\alpha &=& -\frac{C_{ij}}{\vec{B} \cdot \vec{\xi}} \xi^i a^j,
\end{eqnarray}
The eigenvalues of the matrix $D$ (\ref{L3.1}) determine the growth and propagation of the infinitesimal 
plane-wave disturbance. 

From (\ref{L3.1}) it can be shown that $D$ has at least one zero eigenvalue $\lambda_1=0$. Since we are 
working in two dimensions, the remaining eigenvalue is equal to the trace of D: $\lambda_2=Tr(D)$. This 
trace is given by
\begin{equation}
\label{L4}
Tr(D)  =  Tr(C)-\frac{C_{ij} \xi^i B^j}{\vec{\xi} \cdot \vec{B}},
\end{equation}
\[ =  - i (\vec{u}\cdot\vec{\xi}) + {\xi^i B^j \over \vec{\xi} \cdot \vec{B}} 
\left(  \nabla_j u_i + A_{ik} \xi^l \nabla^k A_{jl} 
+ A_{ij} \xi_l \left( A^{lk} q_k + i \mu \nabla_k A^{lk} \right) \right).
\]
where we have used the fact that the matrix is symmetric $A_{ij}=A_{(ij)}$, is trace-free $A_{ii}=0$ and 
further that $Tr(A_{ik}A_{kj})=1$.

We now consider the short wavelength ($|\xi | \rightarrow \infty $) limit of (\ref{L4}). If we write 
$\vec{\chi} = \vec{\xi}/ |\xi|$, then we can examine the powers of $|\xi|$ on the right-hand side of (\ref{L4}). 
The leading order term in $\xi$ on the right-hand side goes as $O(|\xi|^2)$ and is real with coefficient
\begin{equation}
\label{L5}
- \mu \frac{(\chi_l A_{lm} \chi_m) (\delta_{ik} - k A_{ik}) \chi^k  A_{ij} \chi^j}
{(\delta_{pq}-kA_{pq})\chi^p \chi^q} 
\end{equation}
This leading order term was considered by Schaeffer (\cite{schaeffer87}) in his analysis of (\ref{I7}) 
and (\ref{I7.1}). The denominator is always positive for angles of friction $\varphi > 0$. The numerator 
vanishes when $\chi$ points in the direction of the velocity characteristics (these lie at $\pm 45^o$ to the 
principal stress directions) or in the direction of the stress characteristics (these lie at angles of $( \pm 
\varphi + 90^o)/2$ to the principal stress directions) of the background solution. To illustrate this, we 
have computed the eigenvalue $\lambda_2$ for a range of wavevectors for the approximate radial solution 
(given in the appendix). The results are displayed in figure \ref{figureL1}. In this contour plot the four 
wedges of unstable directions approach those given by (\ref{L5}) in the short wavelength limit. 

\begin{figure}
\hspace{1cm}\includegraphics[scale=0.6]{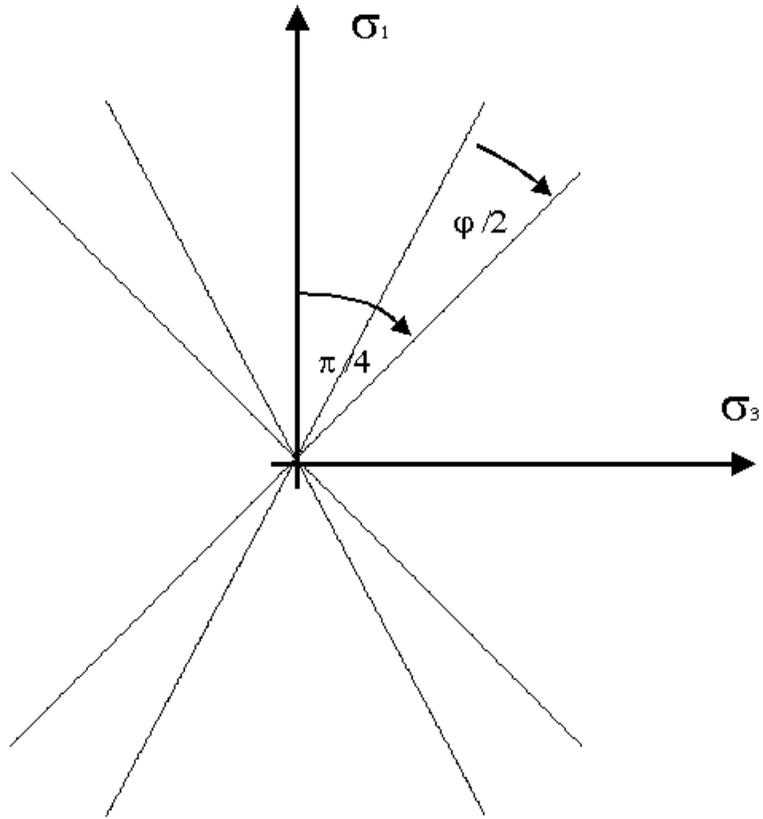}
\caption{Contour plot of the real part of the eigenvalue $\lambda_2$ for the radial solution as a 
function of wavenumber (radial coordinate) and direction of the wave front (angular coordinate). Inside 
the wedges $\lambda_2$ is positive; outside the wedges, $\lambda_2$ is negative.}
\label{figureL1}
\end{figure}

Plane wave disturbances with wavevectors $\xi$ that lie in directions between the stress and 
velocity characteristic directions will be unstable in the short wavelength limit as the real 
part of $\lambda_2$ will be positive in this limit. This short wavelength instability leads us 
to conclude that the two-dimensional granular flow equations are linearly ill-posed (this was 
first noted by Schaeffer \cite{schaeffer87}). 

\section{Singularly perturbed rigid-plastic flow equations}

Attempts to study (\ref{I7}) and (\ref{I7.1}) numerically is problematic. Solutions of these 
equations obtained numerically are found to be grid-dependent due to the wedge of short wavelength
instabilities (\ref{L5}) found in the previous section. Here we propose modifying the equations 
(\ref{I7} - \ref{I7.1}) in an effort to `regularize' the instabilities (\ref{L2}) in some manner. 
We will consider the following equations:
\begin{eqnarray}
\label{P1}
{\partial u^i \over \partial t}+ u^j \nabla_j {u^i} &=& g^i - \nabla_j \left( p \delta^{ij} 
- k p A^{ij}  \right), \\
\label{P1.1}
\nabla_i u^i &=& \epsilon \Delta p.
\end{eqnarray}
These equations are introduced as a tool to study shear-banding rather than an attempt to 
accurately model the physics of dense granular flows. However, it is worth noting some of 
the physical properties of these equations. These equations do not conserve mass as the 
density is held constant despite (\ref{P1.1}). They do satisfy a Drucker-Prager yield 
condition (\ref{I2}) if one assumes a flow rule that is a natural generalization of 
(\ref{I4}):
\begin{equation}
\label{P1.2}
\sigma_{ij} - p \delta_{ij} = \mu \left(V_{ij}- Tr(V) \delta_{ij}/2 \right).
\end{equation}
where $\mu$ is given by (\ref{I5}). Now, in this case
\begin{equation}
\label{P1.3}
k^2 p^2 \geq (\sigma_{ij} - p \delta_{ij})^2 = k^2 p^2 (1- \delta^2/2),
\end{equation}
where $\delta = \epsilon \Delta p / \dnorm{V}$. Note that equation (\ref{P1.1}) guarantees that 
$| \delta | \leq \sqrt{2}$.

If $\delta=0$, then the material is at yield and is incompressible, however as $\Delta p$ increases 
the material relaxes from yield. If $\epsilon$ is sufficiently small, one might expect that in some 
instances the solutions of (\ref{P1}) and (\ref{P1.1}) will approximate smooth solutions of 
(\ref{L2}-\ref{L2.1}). Nonetheless, we will see the inclusion of this term significantly alters the 
behavior of the linearized perturbation equations.

The equations for plane-wave disturbances analogous to (\ref{L2}-\ref{L2.1}) are:
\begin{eqnarray}
\label{P2}
\lambda a_i &=& B_i \alpha + C_{ij} a^{j}, \\
\label{P2.1}
\xi \cdot a &=& -i \epsilon \xi^2 \alpha.
\end{eqnarray}
where $B$ and $C$ are given by (\ref{L2.2}) and (\ref{L2.3}) respectively. 
Using (\ref{P2}) we can eliminate $\alpha$ directly from (\ref{P2}) to give:
\begin{equation}
\label{P3}
\lambda a_i = D_{ij} a_j = (C_{ij} - i {B_i \xi_j \over \epsilon \xi^2}) a^j. 
\end{equation}
The leading order term on the right-hand side of (\ref{P3}) in the short wavelength limit again goes as 
$O(|\xi|^2)$. This term is found to be 
\begin{equation}
\label{P4}
M_{ij} = - {\mu \xi^2 \over Fr^2} M_{ijkl} \chi^k \chi^l.
\end{equation}

The first step in analyzing the eigenvalues of $D$ is to look at the eigenvalues of this leading order
part $M$. In what follows we will set $Fr=1$, and ignore the inertial terms which do not play a role in 
shear banding. If we write
\begin{equation}
\label{P4.5}
\delta = {\epsilon \Delta p \over \dnorm{V}},
\end{equation}
then the trace of $M$ is given by
\begin{equation}
\label{P5}
Tr(M) = - \Lambda \left( 2 + \delta \sqrt{2-\delta^2} \cos(2 \phi) \right).
\end{equation}
and the determinant of $M$ is
\begin{equation}
\label{P6}
Det(M) = {\Lambda^2 \over 2} \left( \sqrt{2-\delta^2} \cos(2 \phi)+\delta \right)^2 
\end{equation}
where $\phi$ is the angle of the wavevector $\xi$ to the principal stresses (here the stress is defined
by equation (\ref{P1.2})), and 
\[
\Lambda = \left( {\mu \xi^2 \over 2} \right).
\]
Noting that $| \delta | \leq \sqrt{2}$, we see the trace of $M$ is always negative. Further, $Det(M)$
is non-negative (if $|\delta | > \sqrt{2-\delta^2}$ then $Det(M)$ is strictly positive). Hence the 
matrix $M$ has no positive eigenvalues, although one of the eigenvalues will vanish in the directions
$\cos(2 \phi) = -\delta /  \sqrt{2-\delta^2}$ provided $|\delta | <  \sqrt{2-\delta^2}$. For a smooth 
background solution with $| \delta | \ll 1$, $M$ will have a zero eigenvalue in four directions.

In the short wavelength limit, the eigenvalues of $D$ will tend to those of $M$ (as these grow as $|\xi|^2$)
except in the four directions where the determinant of $M$ vanishes. Thus in this limit $D$ will possess 
eigenvalues which are negative except very near these four isolated directions. This is in contrast to the 
earlier situation where equations (\ref{L1}) and (\ref{L1.1}) possessed a wedge of unstable directions in 
the short wavelength limit. The possibility still remains that the equations (\ref{P1}) and (\ref{P1.1}) 
will be ill-posed near the four directions $\cos(2 \phi) = -\delta /  \sqrt{2-\delta^2}$.

\section{Numerical solutions}

In this section we present numerical solutions of the equations (\ref{P1}) and (\ref{P1.1}). These equations 
are solved in a two-dimensional wedge-shaped hopper with frictional boundary conditions on the walls 
($\theta=\theta_w$) and stress-free inflow/outflow conditions on the arcs $r=r_0$ and $r=r_1$. The radial
solution described in the appendix is supplied as an initial condition. These computations have been using a 
finite element toolkit called Unstructured Grids (UG) which is freely distributed by the IWR at the University of 
Heidelberg \cite{bastian97}.  

The equations are discretized using a finite-volume scheme. Finite-volume methods have been successfully used to 
solve the Navier-Stokes equations in a variety of situations (Patankar \cite{patankar}). We have chosen to use a 
control volume finite-element method due to Schneider and Raw (\cite{schneider87}; see also Kariman and Schneider 
\cite{kariman95}). This method utilizes colocated primitive flow variables (i.e. average stress and velocities 
which are discretized on the same grid) so is well suited to the formulation of the incompressible rigid-plastic 
flow equations presented in the previous section. The equations are discretized in time using a first-order 
backwards Euler scheme.

The initial difficulty in solving these equations is the non-linearity in the stress terms on the right-hand side 
of (\ref{P1}). The convective terms in (\ref{P1}) are also non-linear. We will use a Newton linearization of these 
non-linear terms and solve the resulting equations using a Newton iteration. Within each Newton iteration, the
resulting linear problem is solved using a biconjugate gradient method with multigrid preconditioning. 

The second difficulty is the growth of the instabilities and the development of shear bands. Once the instabilities 
develop a small time-step must be used. The shear-banding results in discontinuities developing across the 
wavefronts that grow from the instability. This is dealt with in an ad hoc manner; the grid is designed so
that the discontinuities form parallel to element boundaries. The hopper has been discretized using 
grids of up to 4000 linear quadrangular elements. These fine grids are required to accurately resolve the 
right-hand side of (\ref{P1.1}) when $\epsilon \ll 1$. On sufficiently fine grids, one finds that the wavelength of 
the instability is controlled by the magnitude of $\epsilon$.

\begin{figure}
\hspace{-0.5cm}\includegraphics[scale=.6]{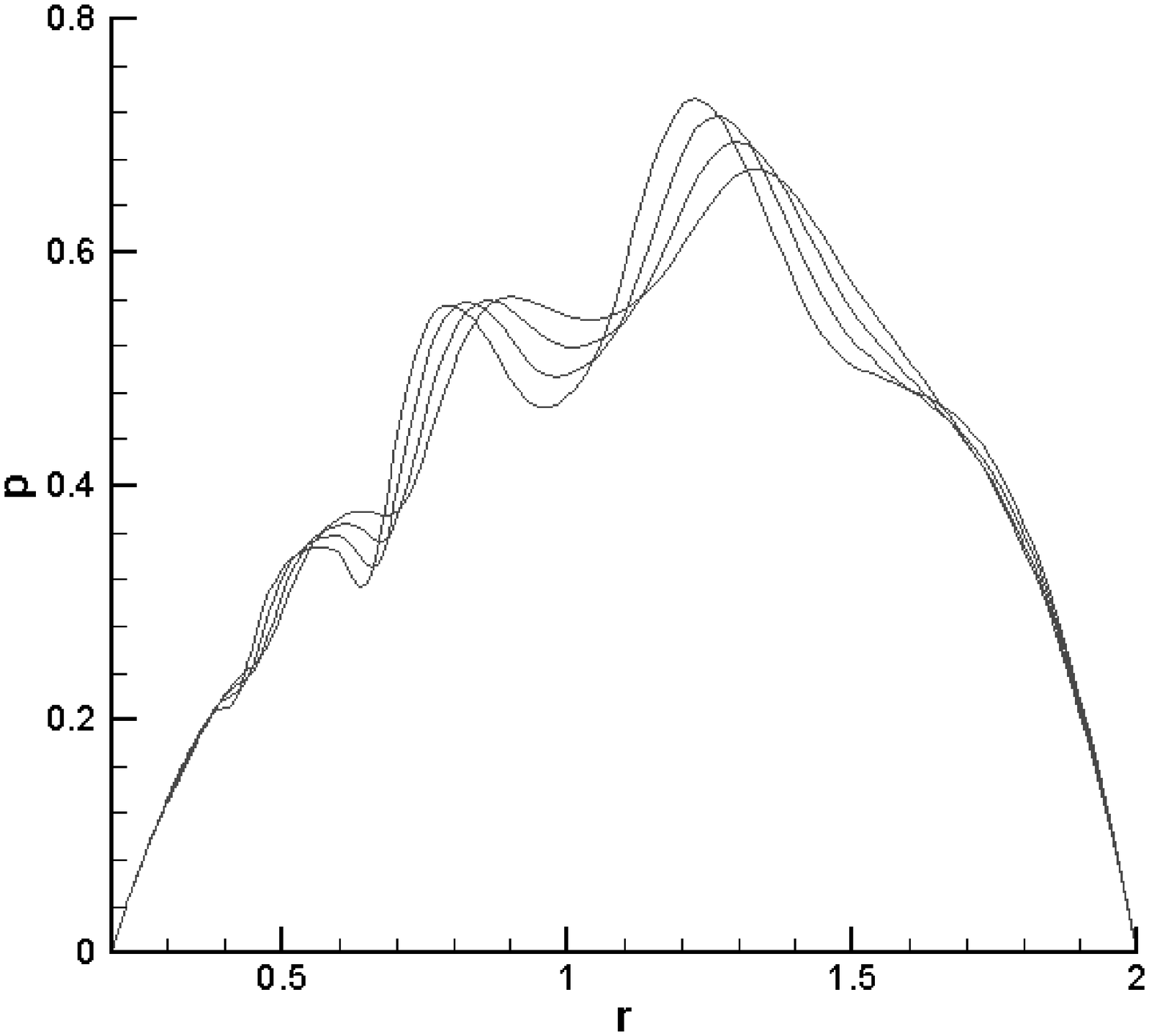}
\caption{Radial slices (constant $\theta$) of the average stress. This shows the wavefronts
propagating down the hopper. Snapshots are taken at 0.025 s intervals.}
\label{figureN3}
\end{figure}

Figures~\ref{figureN1} and \ref{figureN2} depict the evolution of the flow from the smooth radial flow solution
(see appendix). The figures show flows with $\varphi=34^o$ and $\varphi_w=9^o$. The parameter $\epsilon$ is chosen to be 
$2\times10^{-5}$. The flows develop as follows:
\begin{itemize}
\item Initially, plane wave-like wavefronts in the stress develop near the center of the hopper. This region
corresponds to the location in the hopper where $\mu$ is largest. 
\item At these wavefronts, the tangential velocity component begins to develop jumps across the trough of
each wavefront.
\item These instabilities develop further and further down the hopper, and the formation of jumps in
the tangential velocity across the wavefronts follows.
\item The growth of the wavefronts in amplitude does not increase without bound. Eventually, the growth appears
to cut off. The wavefronts are observed to propagate down the hopper (see figure~\ref{figureN3} where 
$\varphi=34^o$ and $\varphi_w=9^o$ but $\epsilon$ is $1\times10^{-4}$ so that the wavelength observed here
is longer than in preceding figures). 
No steady state emerges.
\end{itemize}
In the next section, we will analyze the linear stability of this system in further detail to try to understand 
these features. These features very much resemble shear bands, particularly the jumps in tangential velocity that 
occur across the wavefronts. It should be noted that the thickness across which this jump occurs seems to be 
grid-dependent. Thus, the shear bands are not fully resolved by the numerical method.

\begin{figure}
\begin{center}
\includegraphics[scale=.55]{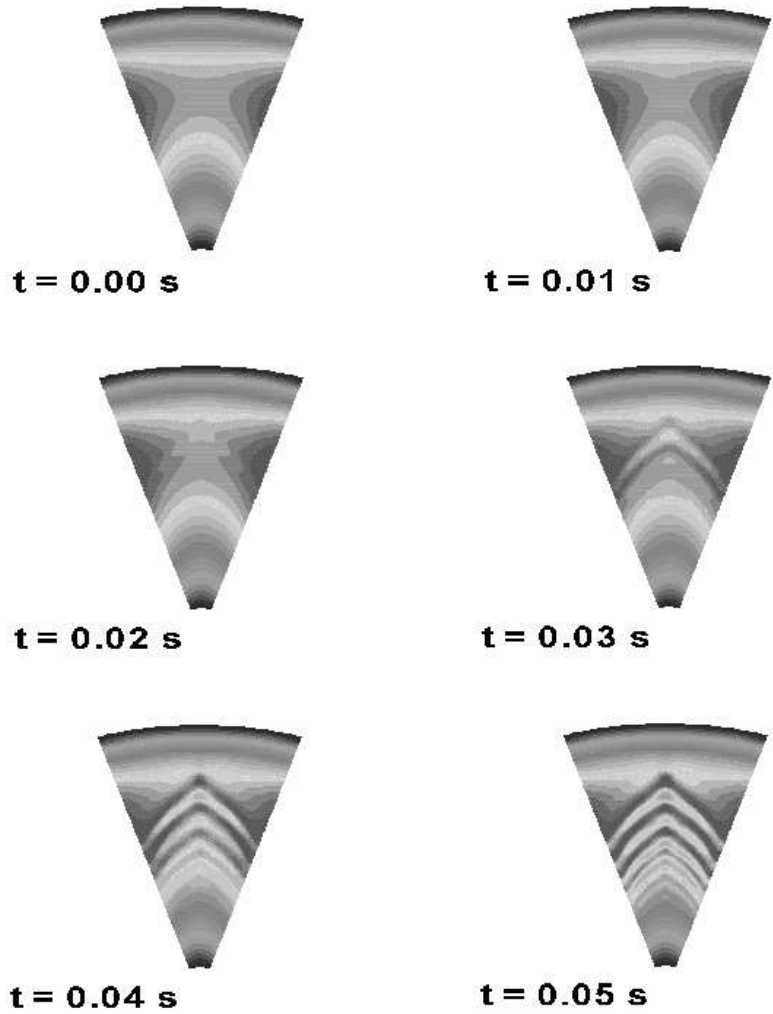}
\end{center}
\caption{Contour plots of the average stress in a wedge-shaped hopper showing the development of
instabilities. Snapshots have been taken at 0.01 s intervals. These instabilities lead to 
shear-banding.}
\label{figureN1}
\end{figure}

\begin{figure}
\hspace{-1cm}\includegraphics[scale=.4]{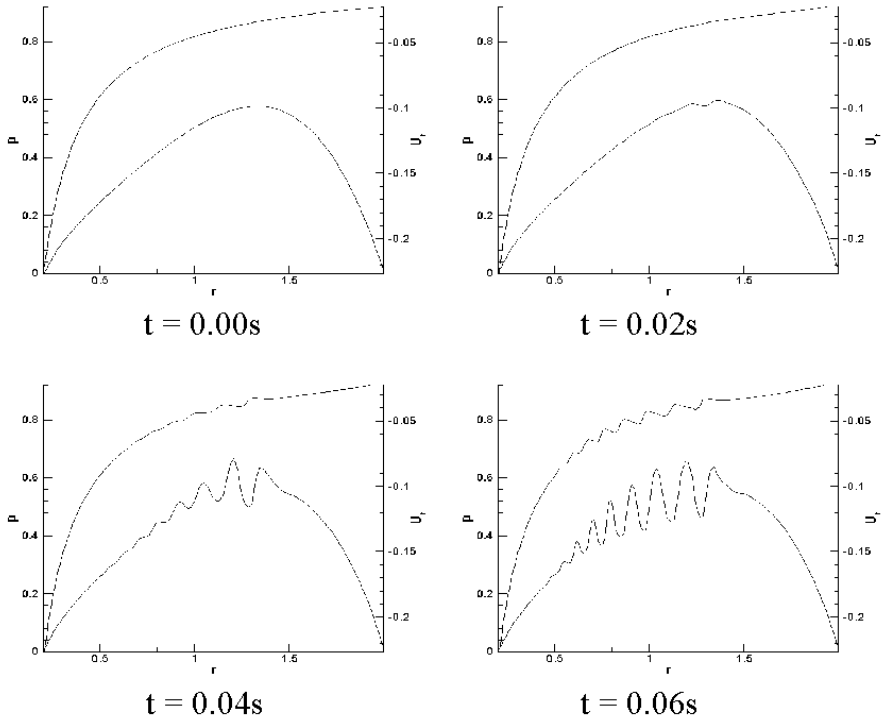}
\caption{Radial slices (constant $\theta$) of the average stress, and the tangential velocity component 
across the band. The normal velocity components (not shown) are continuous across the bands. Snapshots are
taken at 0.02 s intervals.}
\label{figureN2}
\end{figure}

\section{Instabilities in the singularly perturbed system}

Including the singular perturbation in (\ref{P1.1}), introduces a new scale, $\epsilon$, to the problem. The 
grouping $\sqrt{\mu \epsilon}$ gives a length-scale that appears in the modified eigenvalue problem (equation 
\ref{P3}). Indeed, the wavenumber of the instabilities in the hopper interior that are observed in the numerical 
studies above roughly satisfy $\mu |\xi|^2 \sim {1 \over \epsilon} \gg | \xi|$, so this length-scale appears to 
be relevant for these features at least. This suggests that the eigenvalues of the following truncation of the 
full tangent matrix $D$ should be sufficient to study the stability of these unstable modes:
\begin{equation}
\label{P7}
D^\prime_{ij} = M_{ij} + {1 \over \epsilon} \left(k A_{ik} - \delta_{ik} \right) \chi^k \chi_j. 
\end{equation}
This truncation $D^\prime$ neglects the imaginary parts of $D$, and the contribution from the inertial terms. 

The trace and determinant of $D^\prime$ are given by:
\begin{equation}
\label{P8T}  
Tr(D^\prime)= Tr(M) - {1 \over \epsilon}
\left(1+{k \over 2} (\sqrt{2-\delta^2} \cos(2 \phi)-\delta)\right),
\end{equation}
\begin{equation}
\label{P8D}
Det(D^\prime) = Det(M) + {\Lambda \over \epsilon} \left( (2-\delta^2) \cos^2(2 \phi) 
+ k \sqrt{2-\delta^2} \cos(2 \phi) + \delta (\delta -k)  \right).
\end{equation}
In the short wavelength limit $Tr(D^\prime) \sim Tr(M)$ and, unless $Det(M)$ vanishes, 
$Det(D^\prime) \sim Det(M)$. If, however, $Det(M)$ vanishes then 
\begin{equation}
\label{P9}  
Det(D^\prime) \sim {2 \Lambda \over \epsilon} \delta (\delta -k).
\end{equation}
so that the determinant of $D^\prime$ will be negative if $0< \delta < k$. In this case $D^\prime$ 
will possess a positive eigenvalue where $Det(M)$ vanishes. Thus it seems plausible that under certain 
conditions (\ref{P1}) and (\ref{P1.1}) can become ill-posed along the directions where $Det(M)$ 
may vanish. We should be cautious, however, in drawing conclusions about the short wavelength behavior 
of $D$, from the short wavelength behavior of $D^\prime$.

\begin{figure}
\includegraphics[angle=270,scale=.6]{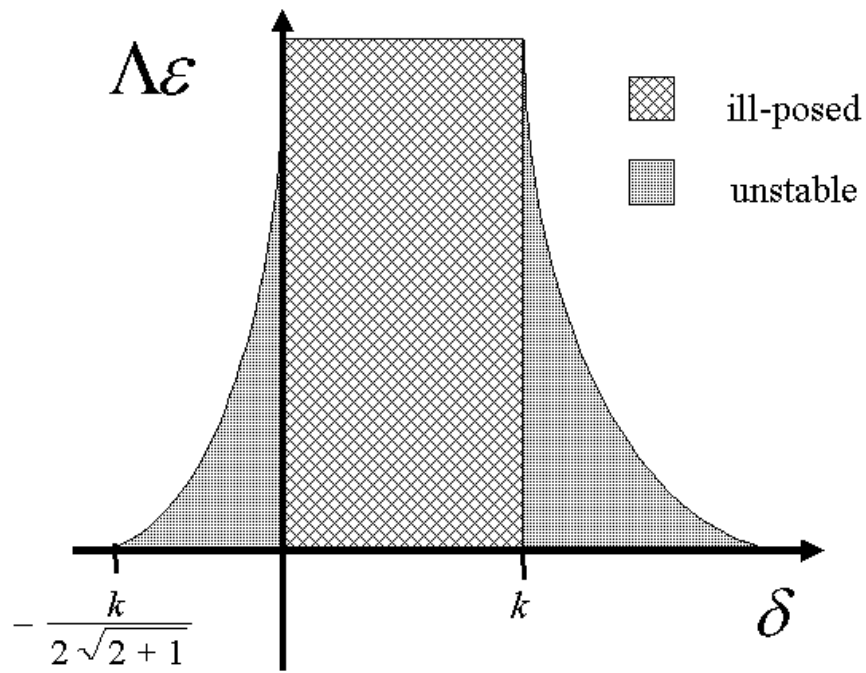}
\caption{Schematic stability phase diagram in terms of the $\Lambda \epsilon$ (which depends on frequency) and the 
parameter $\delta$. There is now dependence on the direction of the wavevector is this plot; for points on the plot
denoted unstable or ill-posed, there exists at least one direction with this property.}
\label{figureP1}
\end{figure}

Away from the short wavelength limit, examination of $Det(D^\prime)$ reveals the possibility of 
other instabilities. The existence of these instabilities depends on the size and sign of $\delta$.
There are three possibilities which are classified below:
\begin{enumerate}
\item $0 \geq \delta > - {k \over 2 (\sqrt{2}+1)}$: $Det(D^\prime)$ is non-negative for all wavenumbers
and all directions.
\item $\delta > k$ or $-{k \over 2 (\sqrt{2}+1)} > \delta$: $Det(D^\prime)$ can be negative for 
wavenumbers that satisfy 
\[
0< \Lambda \epsilon < 2 (k^2-4 \delta (\delta-k))/ \delta (\delta-k).
\]
Thus $Det(D^\prime)$ is non-negative in the short wavelength limit (the problem is linearly well posed).
\item $k \geq \delta < 0$: $Det(D^\prime)$ can be negative at any wavelength (i.e. this suggests the problem 
may be linearly ill-posed).
\end{enumerate}
The situation is illustrated schematically in figure~\ref{figureP1}.

\begin{figure}
\includegraphics[scale=.6]{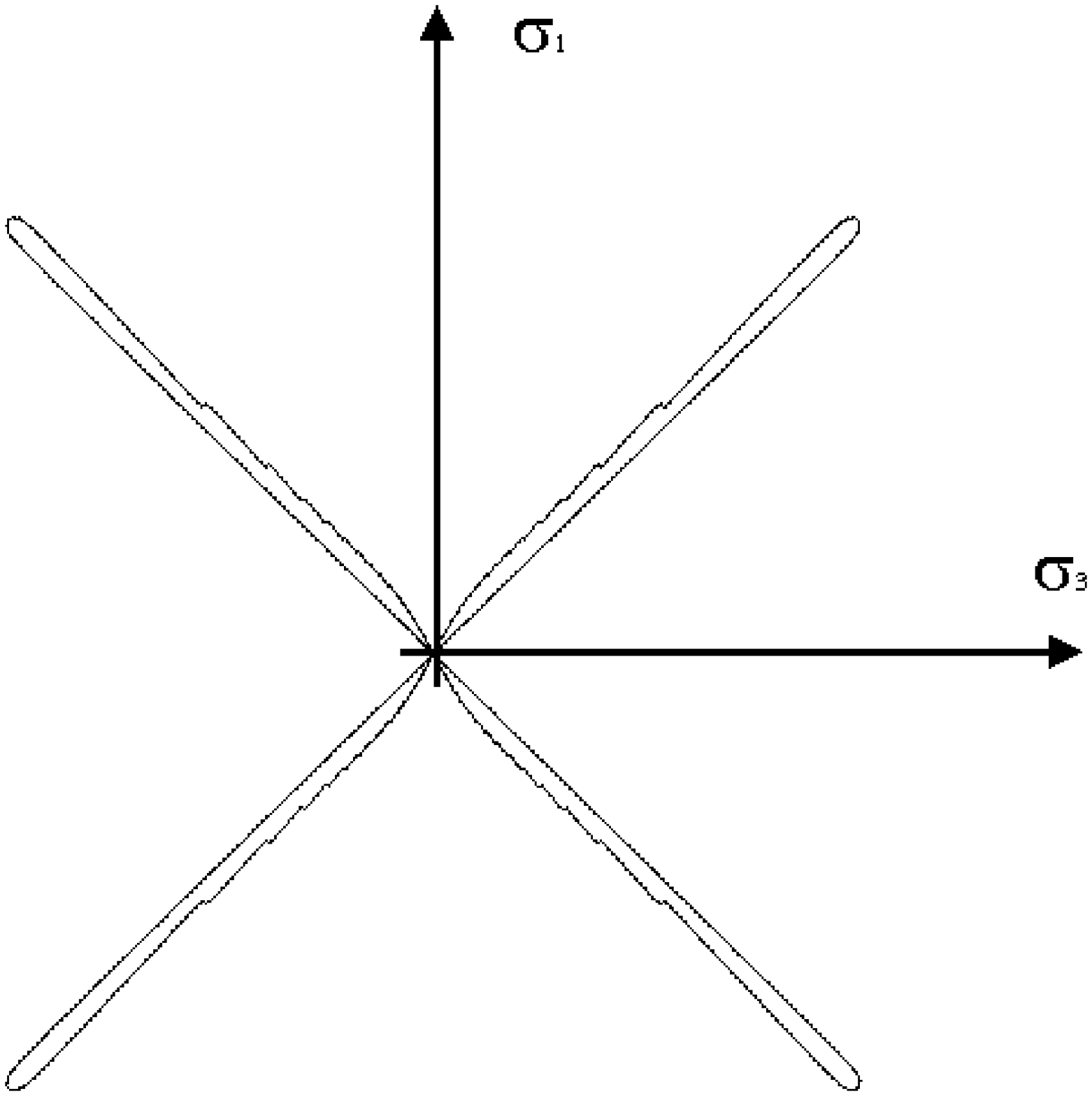}
\caption{Contour plot of the real part of the eigenvalue $\lambda_2$ for the radial solution as a 
function of wavenumber (radial coordinate) and direction of the wave front (angular coordinate). 
The contour depicted is the $\lambda_2=0$ contour. Inside the contour $\lambda_2$ is positive; outside 
this contour, $\lambda_2$ is negative. Here we can see the unstable region disappears as $|\xi| 
\rightarrow \infty$.}
\label{figureP2.1}
\end{figure}

This provides a qualitative explanation of the development of shear bands observed in the numerical 
simulations detailed in the previous section. The initial smooth background solution will have small, 
negative values of $\delta$, so the problem is initially well-posed everywhere but unstable to perturbations 
of wavenumber $|\xi | \sim 1/ \sqrt{\mu \epsilon}$. As an example, the determinant of the full tangent matrix 
$D$ has been computed for the radial background solution near the center of the hopper (see appendix), with 
$\varphi=34^o$, $\varphi_w=9^o$ and a value of $\epsilon = 2 \times 10^{-5}$. Illustrated in 
figure~\ref{figureP2.1}, we can see that the unstable region disappears as $|\xi| \rightarrow \infty$, 
which is consistent with the behavior of $D^\prime$. 

After a short time, an instability does develop with wavenumber $|\xi | \sim 1/ \sqrt{\mu \epsilon}$ 
and wavevector at approximately $45^\circ$ to the directions of principal stress.  Once the disturbance 
grows sufficiently large, equations (\ref{P2}-\ref{P2.1}) will cease to describe the behavior of the 
disturbance. Nonetheless, we can continue a qualitative description of the process using (\ref{P2}-\ref{P2.1}) 
as a guide. As this disturbance grows then, $\delta$ becomes a rapidly changing function of position in the 
hopper. In the troughs of the plane-wave-like pressure disturbance, $\delta$ is positive, and the linear 
stability analysis would suggest that the solution becomes ill-posed in this case. Comparing this with the 
numerical simulations suggests that this ill-posedness in the troughs of the pressure disturbance causes 
a very strong gradient in the velocities to develop, which in turn drives $\delta$ to zero. This feedback 
may provide a mechanism to regularize the ill-posedness. This needs to be studied further, by looking at 
length scales other than $\sqrt{\mu \epsilon}$ and by utilizing a non-linear analysis more suited to dealing 
with the inhomogeneities that develop as the instabilities grow. 

\section{Discussion}

We have presented new equations which exhibit shear-banding behavior. These equations are related to 
the IRPF equations by a singular perturbation. This singular perturbation regularizes the linear 
ill-posedness inherent in the IRPF equations by collapsing the wedge of wave directions that are unstable
in the short wavelength limit. These resulting equations still possess instabilities, related to those of the 
IPRF as seen in the linear stability analysis, and in numerical simulation these instabilities lead to 
shear-banding. The linear stability analysis suggests that the inhomogeneity that develops with these 
instabilities leads to further short wavelength instabilities, which cause these shear-band like features 
to develop. 
 
The wavelength of the instabilities observed in the singularly perturbed equations was controlled by the 
time-scale $\epsilon$. The thickness across the jumps in tangential velocity across the wavefronts was 
found to be grid-dependent however. The linear stability analysis considered here was concerned with length 
scales $\sqrt{\mu \epsilon}$ and did not predict a thickness across these jumps. However the analysis suggests 
that they are associated with the ill-posedness that can develop under certain circumstances. It is possible 
that the thickness of the shear bands here is zero; in this case, it would be impossible to fully resolve the 
band numerically. Indeed, we were not able to resolve the bands in our numerical simulation. However, the linear 
analysis does suggest a mechanism whereby the ill-posedness may be shut-off as large velocity gradients appear 
in ill-posed regions. 

The stability properties of these equations is similar to that of the critical state plasticity studied by Schaeffer 
\cite{schaeffer90}. Indeed, the parameter $\delta$ (equation \ref{P4.5}) here plays a role similar to the dilation 
angle in CST. However, the equations proposed here are considerably simpler than the CST equations and may be a 
simpler system in which to study shear banding (the CST equations, however, will certainly provide a better model 
for granular flows). For instance, it may be possible to study the non-linear stability of this system to obtain a 
deeper understanding of the development of shear bands and their relationship to ill-posed behaviour. 

\section*{Acknowledgments}
The author thanks Dr Graham Weir and Dr Roger Young for their help and guidance. 
\appendix

\section{Approximate radial solutions}

For wedge-shaped hoppers Brennen and Pearce \cite{brennen} and Kaza and Jackson \cite{jackson82}, 
have examined steady-state plane-strain flows by expanding the velocities and the pressure in powers 
of $(\theta/ \theta_w)$ where $\theta_w$ is the wall-angle of the wedge-shaped hopper. The steady-state 
equations are then solved order by order using a radial-flow ansatz. The first order terms in these 
expansions are similar to the radial solutions of Jenike for the quasistatic ($Fr \ll 1$) steady-state 
equations (\ref{I7}) and (\ref{I7.1}). These approximations contain corrections due to the inclusion 
of the inertial terms when $Fr \sim 1$. Working in polar coordinates ($r,\theta$ where the hopper walls 
lie at $\pm \theta_w$) in the plane and using the Sokolovski parameterization of the stresses 
(\cite{sokolovski}), the ansatz goes as follows:  
\begin{eqnarray}
p & = & p_0(r) + O  \left( {\theta\,^2 \over {\theta_w}^2} \right), \\
\label{R2} \Psi & = & \Psi_1 \left( {\theta \over \theta_w} \right) + O  \left( {\theta\,^3 \over {\theta_w}^3} \right),
\end{eqnarray}
where $\Psi$ is related to the stresses by
\[
\Psi = {2 \sigma_{r \theta} \over \sigma_{rr}-\sigma_{\theta \theta}}. 
\]
The leading order terms in $\theta$ are then given by:
\begin{eqnarray}
\label{R7} p_0(r) &=&  { 5_0 \over 1-\sin\varphi} \left[ \left( {F^2 \over w+2} - {1 \over w-1} \right) 
\left( r \over r_0 \right)^w +{r/r_0 \over w-1} -{F^2 \over w+2 } \left({r_0 \over r} \right)^2 \right] \\
\label{R8} \Psi_1 &=& \Psi_w,
\end{eqnarray}
where $\Psi_w$ is the value of $\Psi$ at the wall (determined by the boundary condition (\ref{I6})), 
\[
w= { 2 \sin \varphi \over (1-\sin \varphi)} \left( 1 + \Psi_w/ \theta_w \right). 
\]
and
\[
F^2= \left( {w+2 \over w-1} \right) \left({1- \left({r_0 \over r_1}\right)^{w-1} 
\over 1- \left({r_0 \over r_1}\right)^{w+2} } \right).
\]
Here $r=r_0$ and $r=r_1$ are the arcs along which the average stress vanishes (the outflow and inflow respectively).
The corresponding terms in the velocity fields are also radial (to order $\left(\theta/ \theta_w\right)^2$): 
\begin{eqnarray}
\label{R9} u & = &  \sqrt{g r_0} F \left(1-2 \theta_w \Psi_w \left({\theta \over \theta_w}\right)^2 \right) 
{r_0 \over r} \\
\label{R10} v & = & 0,  
\end{eqnarray}
where $u$ and $v$ are the radial and angular components of the velocity respectively.


\begin{thebibliography}{10}

\bibitem{bastian97}
P.~Bastian, K.~Birken, K.~Johannsen, S.~Lang, K.~Eckstein, N.~Neuss,
  H.~Rentz-Reichert, and C.~Wieners.
\newblock Ug - a flexible software toolbox for solving partial differential
  equations.
\newblock {\em Computing and Visualization in Science}, 1:27--40, 1997.

\bibitem{brennen}
C.~Brennen and J.~C. Pearce.
\newblock Granular material flow in two-dimensional hoppers.
\newblock {\em Journal of Applied Mechanics}, 45:43, 1978.

\bibitem{drescher}
A.~Drescher.
\newblock On the criteria for mass flow in hoppers.
\newblock {\em Powder Technology}, 73:251--260, 1992.

\bibitem{drucker}
D.~C. Drucker and W.~Prager.
\newblock {\em Q. Appl. Math.}, 10:157, 1952.

\bibitem{gremaud99}
P.-A. Gremaud and J.~V. Matthews.
\newblock On the computation of steady hopper flows: I, stress determination
  for coulomb materials.
\newblock Technical Report CRSC TR99-35, Center for Research in Scientific
  Computing, North Carolina State University, Raleigh, 1999.

\bibitem{jenike}
A.~Jenike.
\newblock Gravity flow of bulk solids.
\newblock Technical report, Utah Engineering Experimental Station, University
  of Utah, Salt Lake City, 1964.

\bibitem{kariman95}
S.~M.~H. Kariman and G.~E. Schneider.
\newblock Pressure-based control-volume finite element method for flow at all
  speeds.
\newblock {\em AIAA Journal}, 33(9):1611, 1995.

\bibitem{jackson82}
K.~R. Kaza and R.~Jackson.
\newblock The rate of discharge of coarse granular material from a wedge-shaped
  mass flow hopper.
\newblock {\em Powder Technology}, 33:223, 1982.

\bibitem{muhlhaus96}
H.-B. Muhlhaus and F.~Oka.
\newblock Dispersion and wave propagation in discrete and continuous models for
  granular materials.
\newblock {\em Journal of Solids and Structures}, 33:2841--2858, 1996.

\bibitem{nedderman92}
R.~M. Nedderman.
\newblock {\em Statics and Kinematics of Granular Materials}.
\newblock Cambridge University Press, London, 1992.

\bibitem{patankar}
S.~V. Patankar.
\newblock {\em Numerical Heat Transfer and Fluid Flow}.
\newblock Hemisphere, Washington D.C., 1980.

\bibitem{schaeffer87}
D.~G. Schaeffer.
\newblock Instability in the evolution equations describing incompressible
  granular flow.
\newblock {\em Journal of Differential Equations}, 66:19, 1987.

\bibitem{schaeffer90}
D.~G. Schaeffer.
\newblock Instability and ill-posedness in the deformation of granular
  materials.
\newblock {\em International Journal for Numerical and Analytical Methods in
  Geomechanics}, 14:253--278, 1990.

\bibitem{schaeffer92b}
D.~G. Schaeffer.
\newblock A mathematical model for localization in granular flow.
\newblock {\em Proceedings of the Royal Society of London A}, 436:217--250,
  1992.

\bibitem{schneider87}
G.~E. Schneider and M.~J. Raw.
\newblock Control volume finite-element method for heat transfer and fluid flow
  using colocated variables - 1. computational procedure.
\newblock {\em Numerical Heat Transfer}, 11:363, 1987.

\bibitem{sokolovski}
V.~V. Sokolovski.
\newblock {\em Statics of Granular Materials}.
\newblock Pergamon, Oxford, 1965.

\bibitem{thorpe92}
R.~B. Thorpe.
\newblock An experimental clue to the importance of dialtion in determining the
  flow rate of a granular material from a hopper or bin.
\newblock {\em Chemical Engineering Science}, 47:4295--4303, 1992.

\bibitem{peters96}
K.~C. Valanis and J.~F. Peters.
\newblock Ill-posedness of the initial and boundary value problems in
  non-associative plasticity.
\newblock {\em Acta Mechanica}, 114:1--25, 1996.

\end{thebibliography}
\end{document}